\documentclass[fleqn,usenatbib]{mnras}

\usepackage{newtxtext,newtxmath}

\usepackage[T1]{fontenc}
\usepackage{ae,aecompl}

\usepackage{graphicx}	
\usepackage{amsmath}	
\usepackage{amssymb}	
\usepackage{threeparttable}
\usepackage{url}

\title[Prompt radio emission during GRB X-ray flares]{LOFAR detectability of prompt low-frequency radio emission during gamma-ray burst X-ray flares}

\author[Starling et al.]{
R. L. C. Starling,$^{1}$\thanks{ASTRON Helena Kluyver visiting fellow. E-mail: rlcs1@le.ac.uk} A. Rowlinson$^{2,3}$, A. J. van der Horst$^{4,5}$ and R. A. M. J. Wijers$^{3}$\\
$^{1}$ School of Physics and Astronomy, University of Leicester, University Road, Leicester LE1 7RH, UK\\
$^{2}$ Netherlands Institute for Radio Astronomy (ASTRON), PO Box 2, 7990 AA Dwingeloo, The Netherlands\\
$^{3}$ Anton Pannekoek Institute, University of Amsterdam, Postbus 94249, 1090 GE Amsterdam, The Netherlands\\
$^{4}$ Department of Physics, The George Washington University, 725 21st St. NW, Washington, DC 20052, USA\\
$^{5}$ Astronomy, Physics and Statistics Institute of Sciences (APSIS), The George Washington University, Washington, DC 20052, USA
}

\date{Accepted XXX. Received YYY; in original form ZZZ}

\pubyear{2020}

\begin{document}
\label{firstpage}
\pagerange{\pageref{firstpage}--\pageref{lastpage}}
\maketitle

\begin{abstract}
The prompt emission in long gamma-ray bursts arises from within relativistic outflows created during the collapse of massive stars, and the mechanism by which radiation is produced may be either magnetically- or matter-dominated. In this work we suggest an observational test of a magnetically-dominated Poynting flux model that predicts both $\gamma$-ray and low-frequency radio pulses. A common feature among early light curves of long gamma-ray bursts are X-ray flares, which have been shown to arise from sites internal to the jet. Ascribing these events to the prompt emission, we take an established {\it Swift} XRT flare sample and apply a magnetically-dominated wind model to make predictions for the timing and flux density of corresponding radio pulses in the $\sim$100--200\,MHz band observable with radio facilities such as LOFAR. We find that 44 per cent of the X-ray flares studied would have had detectable radio emission under this model, for typical sensitivities reached using LOFAR's rapid response mode and assuming negligible absorption and scattering effects in the interstellar and intergalactic medium. We estimate the rate of {\it Swift} gamma-ray bursts displaying X-ray flares with detectable radio pulses, accessible to LOFAR, of order seven per year. We determine that LOFAR triggered observations can play a key role in establishing the long debated mechanism responsible for gamma-ray burst prompt emission.
\end{abstract}




\section{Introduction} \label{sec:intro}

Gamma-ray bursts (GRBs) show a rich diversity in their temporal behaviour, particularly in the early stages of their evolution. The prompt emission is usually recorded by $\gamma$-ray instruments operating at energies above $\sim$10\,keV, and consists of one or more pulses with varying peak fluxes, widths and separations. 
The prompt $\gamma$-ray emission may arise from either a matter-dominated, hydrodynamical flow dissipating energy at shock fronts \citep[e.g.][]{Rees1994} or from a magnetically-dominated outflow in which energy dissipation can occur via magnetic reconnection \citep[e.g.][]{Spruit2001,sironi2015}. Whether the prompt emission arises from a matter-dominated or magnetically-dominated outflow remains a central outstanding question in GRB physics.

Following the prompt $\gamma$-ray emission, longer wavelength observations are possible after a spacecraft slew; the {\it Neil Gehrels Swift Observatory} \citep[hereafter {\it Swift};][]{Gehrels2004} has a typical prompt slew time of just 100\,s.
Some X-ray observations show a continuation of the pulse-like behaviour of the central engine, before transitioning through a steep temporal decay phase to a power law afterglow understood to be the product of synchrotron radiation \citep[e.g.][]{sari1998}. Superposed on the smooth power law decay light curve, late-time pulse-like features can be seen, termed flares. Flares have been detected across the wavelength range from X-rays, through optical, down to radio, but are most common in the X-ray band possibly due to the impressive coverage of GRB X-ray light curves with {\it Swift}. 

{\it Swift} is a GRB-dedicated facility, triggering on 15--350\,keV $\gamma$-rays with the Burst Alert Telescope \citep[BAT,][]{Barthelmy2005} and rapidly following up in the 0.3--10 keV band with the X-Ray Telescope \citep[XRT,][]{Burrows2005}, and in multiple bands spanning 170--650 nm with the UV and optical Telescope \citep[UVOT,][]{Roming2005}. Among all BAT-triggered GRBs, the XRT detects $\sim 95\%$ \citep{Evans2009}, of which about half display X-ray flaring activity \citep[48\%,][]{Swenson2014}. 
In a handful of cases, optical flares are seen during high energy flaring periods \citep[e.g.][note that observational coverage is one factor here]{Cucchiara2011,Vestrand2014,Carrillo2014,Troja2017}. The sharp optical peaks are often attributed to a reverse shock--forward shock combination, according to the standard fireball model \citep{sari1999}.

An internal origin for X-ray flares has long been favoured \citep[e.g.][]{Chincarini2007,Falcone2007,Chincarini2010,Bernadini2011}, based on the similarity of their temporal properties to prompt GRB pulses, including relative rise time, duration and decay time and distribution of waiting times \citep{Guidorzi2015}. Both X-ray flares and prompt pulses show a wide variation in both time since trigger and peak flux, and an increased width and decreased peak flux for later time flares \citep[e.g.][]{Guidorzi2015}. 

A connection with the prompt emission is also evident from coincident X-ray flare activity and tens to hundreds of keV $\gamma$-ray pulses \citep[demonstrated in e.g.][currently limited to a small number of GRBs since instrumental coverage in $\gamma$-rays and X-rays must overlap]{Hu2014,Oganesyan2018}. A three year study using data from {\it Swift} and {\it Fermi} \citep{meegan2009} concluded that X-ray flares can be produced via the late internal shock model of prompt emission, and therefore can be effective probes of the central engine and its activity (\citealt{Troja2015}, as could the rarer prompt optical flares, see e.g. \citealt{Troja2017}).
A single power law is an acceptable fit to $\sim$80\% of the X-ray spectra of {\it Swift} GRBs for which BAT and XRT coverage overlaps \citep{Evans2010}. Band \citep{Band1993}, band-cut \citep{Zheng2012} and cut-off power law models are often a better description of spectra across the $\gamma$-ray band. Simultaneous {\it Fermi} GBM, {\it Swift} BAT and XRT spectral fits have shown that in 62 per cent of cases extending these models down to X-ray energies results in a further, low energy break with spectral indices consistent with a single, broadband synchrotron component in the fast cooling regime \citep[][but see also \citealt{Guiriec2016}]{Oganesyan2018}.

 \citet{Wang2013} and \citet{Yi2016} compared GRB flare properties with those of Solar flares and concluded that they are strikingly similar. This motivates the investigation of magnetically-dominated models for the production of X-ray flares and indeed GRB prompt emission \citep[e.g.][]{Smolsky2000,Uhm2016}, while matter-dominated models also remain viable \citep[e.g.][]{Rees1994,Rees2005}. Some studies of {\it Swift} X-ray flares that consider the steeply decaying segment as arising from off-axis jet angles, referred to as high latitude emission or the curvature effect, have concluded that the emitting region must be undergoing bulk acceleration, requiring additional energy supplied by a significant Poynting flux component \citep{Jia2016,Uhm2016b}. 
Given these connections between X-ray and $\gamma$-ray pulses, and some evidence for a magnetically-powered component, we might reasonably expect to apply prompt emission models also to X-ray flares. For instance the magnetically driven wind model proposed by \citet{Usov2000} includes the generation of early radio emission peaking at low frequencies. Probing GRB prompt emission mechanisms through X-ray flares has the advantage that flaring begins with or after the $\gamma$-ray pulses, and can continue for a significant fraction of both prompt and afterglow phases relaxing the requirements on telescope response times. 

Radio observations have traditionally begun hours to weeks after the GRB, as this is when a radio afterglow, caused by interaction of the forward shock with the circumburst medium, is expected \citep[e.g.][]{Katz1994,chandra2012,anderson2018}. Prompt searches for radio emission have also been carried out, constraining prompt radio emission to lie below several thousand Janskys \citep[e.g][]{dessenne1996,balsano1998}. Alas this is too insensitive to probe many of the proposed prompt radio emission mechanisms. Recent studies have focused on searching for fast radio bursts at GHz frequencies in the first few minutes of a GRB, associated with the plateau phases \citep[][]{Bannister2012}, or a deeper search, $\sim$ Jy sensitivity, for prompt (minute to hours) radio emission from short GRBs \citep{kaplan2016,anderson2018b}. This was followed by a catalogue of $<$1 day follow-up observations at 15.7 GHz with the Arcminute Microkelvin Imager (AMI) which increased the number of radio detections of short and long GRBs \citep{Staley2013,anderson2018}. The new detections spanned 0.3--33 days post-burst, meaning that coverage during the prompt emission itself remains largely unexplored.

Recently, a rapid response mode was implemented for the LOw Frequency ARray \citep[LOFAR, ][]{VanHaarlem2013}. LOFAR operates using either its Low Band Antenna (LBA, 10--90 MHz) or with its High Band Antenna (HBA, 120--240 MHz) and is capable of responding to transient alerts within 5 minutes of receiving the trigger notice. \citet{Rowlinson2019b} outline the first LOFAR search for prompt radio emission in long GRB\,180706A, beginning 4.5 minutes after the $\gamma$-ray trigger. By utilising the full capability of LOFAR, the observational constraints for coherent radio emission can now reach mJy sensitivities, revolutionising the search for this emission from many transient events.

In this paper we explore the detectability of low-frequency radio waves associated with X-ray flares, under the assumptions of a Poynting flux model for the prompt emission mechanism in which X-ray flares are a manifestation of prompt non-thermal pulses. We draw a subset of X-ray flares from an established sample, and apply the model to estimate the fraction of low-frequency waves that would have been detectable by LOFAR for a given set of observational conditions. We discuss the optimal strategy to apply to the current LOFAR triggering and data processing, to find the emission from flares expected within this framework.

\section{Model} \label{sec:model}
Low-frequency waves have been postulated to arise at the shock front of a highly magnetised, relativistic wind (generated by the rotational energy of a compact object) interacting with its surroundings \citep{Smolsky2000,Usov2000}. In this model, a variable current flowing between the two media either side of the travelling shock front leads to the generation of low-frequency radio waves in the MHz regime.

The radio emission should occur at a peak frequency, $\nu_{max}$, in the observed frame of
\begin{equation} \label{eqn:numax}
    \nu_{max} = \frac{1}{1+z} \frac{B_0}{10^2} ~{\rm MHz}
\end{equation}
where $z$ is the redshift of the GRB and we adopt an observer frame value of $B_0 = 10^2 \epsilon_B^{1/2}$ G for the magnetic field in the wind at the radius of low frequency wave generation \citep{Usov2000}, where $\epsilon_B$ is the fraction of the total energy in the magnetic field. We note there are several assumptions in embedded within Equation \ref{eqn:numax}, such as the properties of the magnetic wind and the deceleration radius \citep[see ][ for further details regarding these assumptions]{Usov2000}.

A delay is expected between the emission of the radio pulse and observation, due to dispersion along the line-of-sight, of
\begin{equation} \label{eqn:pulsedelay}
    \tau(\nu) \sim \frac{DM}{241~\nu^2} ~{\rm s}
\end{equation}
in the observer frame, where $\nu$ is the observer frequency in GHz \citep{Taylor1993}. The dispersion measure, $DM = \int n_e D_L$ pc cm$^{-2}$, can be estimated using the relation derived from intergalactic medium modelling: $DM \sim 1200 z$ \citep[e.g.][]{ioka2003,inoue2004,lorimer2007}. 

The observed duration of the pulse, $\tau_r$, is 
\begin{equation} \label{eqn:pulseduration}
    \tau_r (\nu, \Delta \nu) \sim 2 \frac{\Delta\nu}{\nu} \tau(\nu) ~{\rm s}
\end{equation}
\citep{Usov2000} where $\Delta\nu$ is the bandwidth and $\tau(\nu)$ is the dispersion-induced signal delay in seconds. 

Following \citet{Usov2000} we calculate the expected flux density in the MHz bands by assuming a
ratio, $\delta$, between fluence spectral densities in radio and in $\gamma$-rays.
In the two regimes of non-dispersion-limited and dispersion-limited, the predicted radio flux density, $F_{\nu}$, is 

\begin{equation} \label{eqn:radioflux}
    F_{\nu} = \left\{ \begin{array}{ll} 
    \frac{\delta (\beta - 1)}{\tau_r \nu_{\rm max}} \left( \frac{\nu}{\nu_{max}} \right)^{-\beta} ~\frac{\Phi_{\gamma}}{10^{-23}} ~\rm Jy, & \tau_r \le \frac{2 \Delta \nu}{\nu} \tau(\nu)\\ 
    \frac{\delta (\beta - 1)}{2 \Delta \nu \tau} \left( \frac{\nu}{\nu_{max}} \right)^{1-\beta} ~\frac{\Phi_{\gamma}}{10^{-23}} ~\rm Jy, & \tau_r > \frac{2 \Delta \nu}{\nu} \tau(\nu)\\ 
    \end{array} \right.
\end{equation}
where $\delta \sim 0.1 \epsilon_{\rm B}$,~$\beta$ is the power law spectral slope in the high-frequency tail typically assumed to be 1.6, $\tau_r$ is the intrinsic duration of the radio pulse (s), $\Delta \nu$ is the bandwidth (Hz), $\tau(\nu)$ is the dispersion-induced signal delay (s) and $\Phi_{\gamma}$ is the bolometric $\gamma$-ray fluence spectral density (erg cm$^{-2}$). The fraction of energy in the magnetic field $\epsilon_B = 10^{-3}$ is adopted following the limits given in \citet{Katz1997}.

Assuming that the radio pulse has the same duration as the X-ray pulse, we determine within which of the two regimes each X-ray flare occurs. The appropriate form of Equation \ref{eqn:radioflux} can then be applied.

The flux density predictions are made for each X-ray flare individually using the above equations. Multiple flares may well occur in an observation window and in that case we could input their combined fluence here in the way that the model is utilised for prompt emission inputs in \citet{Usov2000}.

We note that these predictions are dependent upon the low-frequency radio emission being able to propagate freely through the ambient medium of the gamma-ray burst, the interstellar medium in the host galaxy and the intergalactic medium. Absorption and scattering effects may lead to reduced observable flux densities \citep[for further discussion of these propagation effects see][ and references therein]{Rowlinson2019}.

\section{X-ray flare sample} \label{sec:sample}
We adopt the X-ray flare sample of \citet{Yi2016}, who identified 468 flares occurring during ten years of {\it Swift} operations up to 2015 March, and fitted their temporal evolution using a smoothly broken power law. Requiring a known redshift, we cut their sample to 200 flares across 81 different GRBs spanning $0.257<z<8.23$. Most of our final sample are long GRBs. Two sources have $T_{\rm 90} \le 2$ s (GRBs 070724, 131004); flare parameters are observed to be similar for both short and long GRBs \citep[e.g. ][]{Margutti2011} so, despite the very different progenitors that are inferred, the prompt emission mechanisms may be the same and hence we keep these in our selection. Four sources have no $T_{\rm 90}$ measurement but on inspection these appear to represent the potential new category of ultralong GRBs for which the emission continues into orbit gaps and can only be estimated as a lower limit \citep[GRBs 111209A, 121027A, 121229A and 130925, ][]{Levan2014,Evans2014}. The ultralong GRBs are particularly interesting given the higher fraction of prompt emission occurring after XRT slew and ground-based response times compared with classical GRBs.

The mean observation start time for XRT following a slew is 113.2\,s for this GRB sample and spans 43.9--577.5\,s. If the X-ray flare peaks before the spacecraft slew, it cannot be identified as such, and we note that this introduces a bias in the times we adopt for this study of radio pulses. Our minimum observed (restframe) flare peak time is 59.7\,s (13.1\,s), with a sample mean of 9760\,s (5206\,s).

\begin{figure}
    \centering
    \includegraphics[angle=0,width=0.45\textwidth]{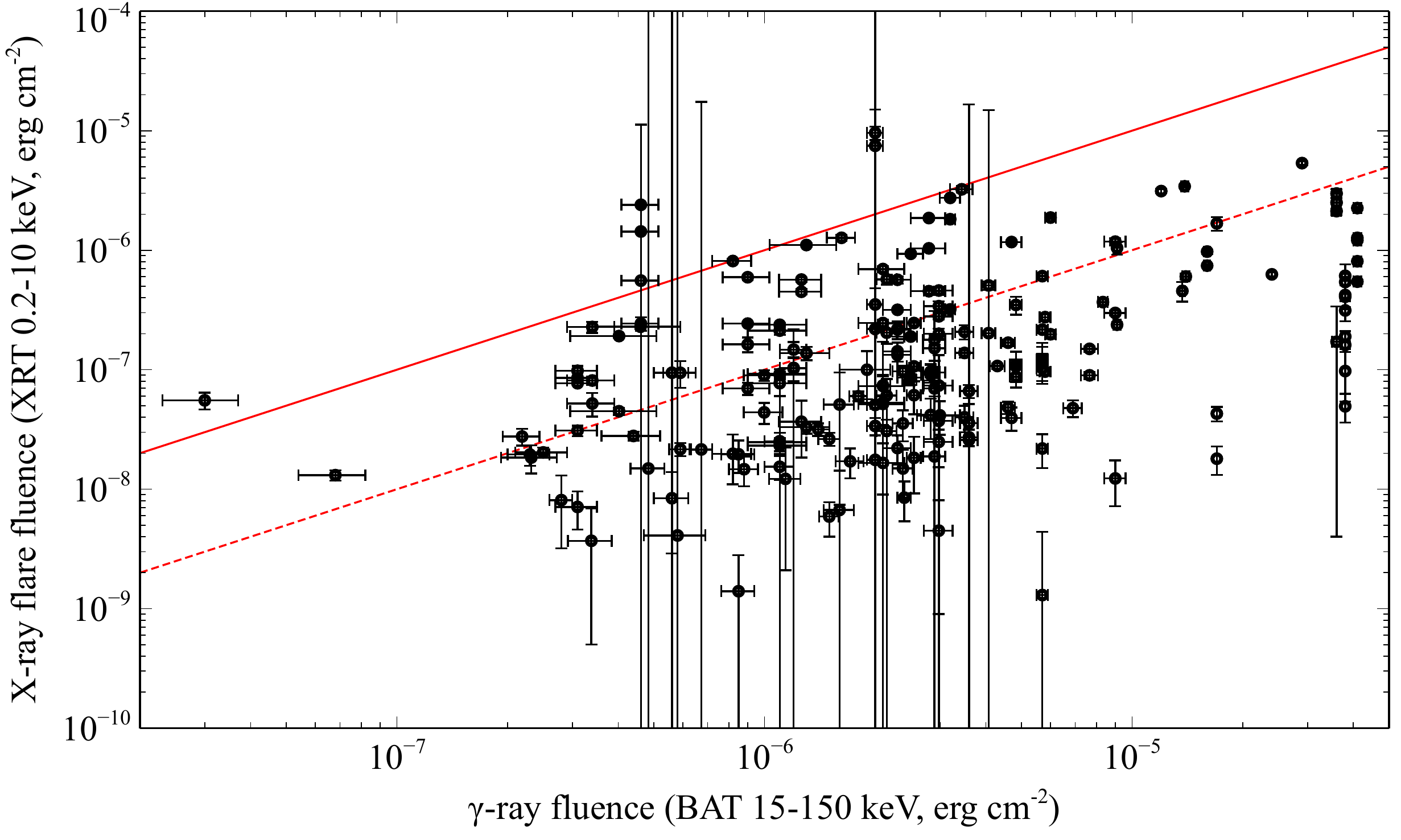}
    \caption{Comparison of the $\gamma$-ray fluence integrated over all observed pulses with the X-ray fluences of individual flares. Errors are 90\% confidence. For reference, red lines indicate the equality line (solid) and 10\% fraction (dashed).}
    \label{fig:fluence}
\end{figure}

The fluence in X-ray flares might be expected to be of order 10 per cent of the $\gamma$-ray fluence \citep[e.g.][]{Chincarini2010}, but rather than making this assumption we use measured X-ray fluences from \citet{Yi2016}. We compare these with the measured $\gamma$-ray fluences from the Third Swift BAT Catalog \citep{Lien2016} in Figure \ref{fig:fluence}\footnote{\url{https://swift.gsfc.nasa.gov/archive/grb_table/}}. We find that the weighted mean fluence ratio of each individual X-ray flare to the total of its GRB in $\gamma$-rays is 0.9$\pm$0.1\% for this sample. 
However, the fluence ratio spans a huge range, from 0.02 to 520 per cent (three GRBs have flares whose X-ray fluence exceeds the total measurable $\gamma$-ray fluence: 060124, 070724A and 121027A), and hence we utilise the high energy fluence for each flare individually in our radio flux density predictions.

\section{Results} \label{sec:results}

Taking the sample of X-ray flares as defined in Section \ref{sec:sample}, we can predict the flux density of the radio flare using Equation \ref{eqn:radioflux} assuming that we trigger a rapid response observation by LOFAR. This method is also directly applicable to observations with other radio facilities, but LOFAR combines both a sufficiently rapid response mode and the deep observational sensitivity required to test this model.

\begin{figure}
    \centering
    \includegraphics[angle=0,width=0.3\textwidth]{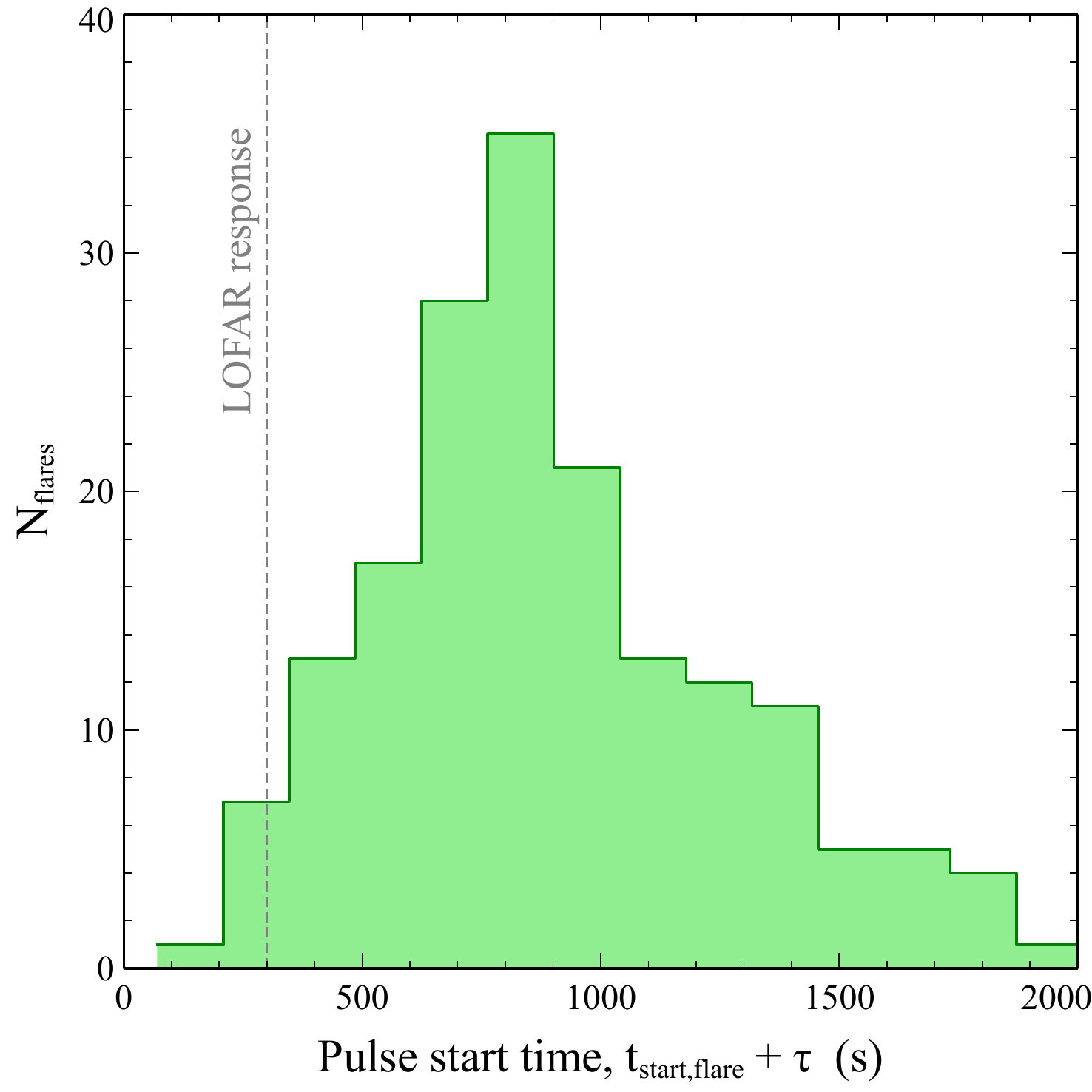}
    \caption{Distribution of expected start times of the radio pulses that occur up to 2000\,s. The current LOFAR response time of 5 minutes is indicated. We note there is a long, low tail to late times but have excluded this here for clarity.} 
    \label{fig:144mhz_delayhist}
\end{figure}
We find the delay time with respect to the X-ray pulse, $\tau$, and duration of the radio flare, $\tau_r$, predicted for each of the X-ray flares in our sample, and plot their distributions in Figure \ref{fig:144mhz_delayhist}. We find mean observed (restframe) values of $t_{\rm flarestart} + \tau = 8758$~(4630) s and $\tau_r = 15523$~(7745)\,s. The sample peak frequencies are $0.003 < \nu_{max} < 0.03$\,MHz (using Equation \ref{eqn:numax}), so currently detectable fluxes will arise from the high-frequency spectral tail.

\begin{figure}
    \centering
    \includegraphics[angle=0,width=0.45\textwidth]{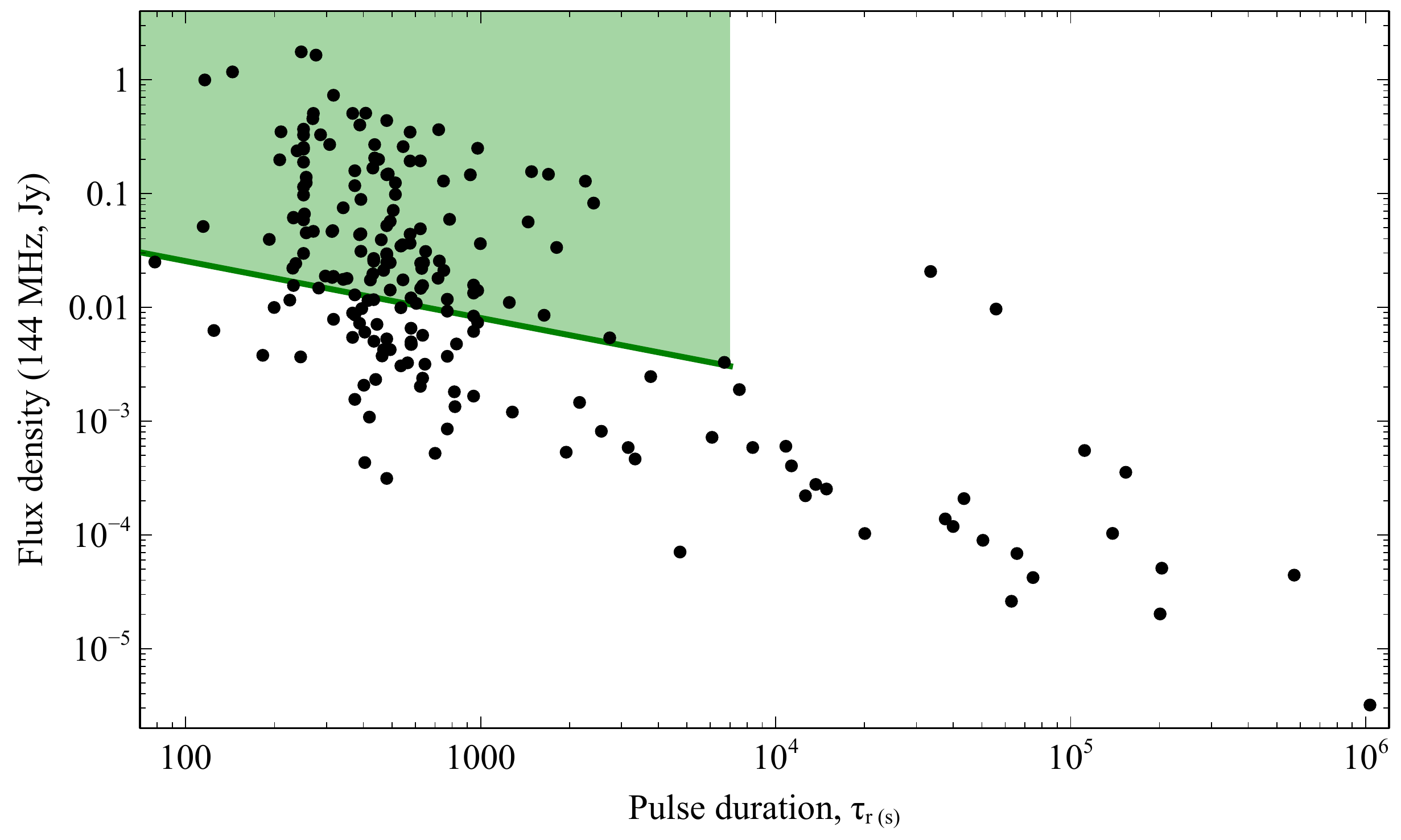}
    \caption{Expectations for flux density at 144 MHz and pulse duration (observer frame). The shaded region indicates detections at or above a 3$\sigma$ level of duration 2 hours or less, based on the sensitivity derived using the rms noise parameters of GRB\,180706 (Equation  \ref{eqn:rmsnoise}).}
    \label{fig:144mhz_flux_duration}
\end{figure}

We calculate the expected flux density in a typical LOFAR HBA GRB observation, following the instrumental setup currently used for rapid response mode LOFAR follow-up of GRBs ($\nu=144$\,MHz, $\Delta \nu = 48$\,MHz, \citealt{Rowlinson2019b}). The results are plotted in Figure \ref{fig:144mhz_flux_duration}.

We can use the triggered observations of GRB\,180706A to quantify the expected sensitivity of future LOFAR observations that are triggered on GRBs containing X-ray flares. The 2 hour integrated image of the field of GRB\,180706A had an RMS noise of $0.6$\,mJy\,beam$^{-1}$ at the location of the GRB, corresponding to a sensitivity of 3\,mJy assuming a $5\sigma$ detection threshold. Although it is not possible to predict the X-ray flares in advance and hence adapt triggering criteria, we are able to tailor our processing of the radio observations to target the flare specifically following the trigger. For instance, a radio flare that is 5 minutes long may be undetectable in a 2 hour integrated image so instead we can produce 5 minute snapshot images in which the background is greatly reduced for shorter pulses \citep[see e.g.][]{Carbone2017,Rowlinson2019b}. Using the general correlation between the sensitivity of a radio image and the integration time ($S \propto t_{\rm int}^{-\frac{1}{2}}$, where $S$ is the detectable flux density in the image and $t_{\rm int}$ is the image integration time in seconds), we extrapolate from the 2 hour image properties to estimate the detection threshold in shorter duration snapshot images using
\begin{eqnarray} \label{eqn:rmsnoise}
	S = 3~{\rm mJy} ~ \left(\frac{t_{\rm int}}{7200 ~{\rm s}}\right)^{-\frac{1}{2}}.
\end{eqnarray}

A total of 86 flares starting after the 5 minutes LOFAR response time met the sensitivity criterion of Equation \ref{eqn:rmsnoise}. This reduces to 85 (42 per cent) of the 200 flares in our sample, when also requiring a start time within the 2 hour integration window. The detectable parameter space is indicated as a shaded region in Figure \ref{fig:144mhz_flux_duration}. These 85 flares are spread among 45 individual GRBs. This calculation assumes no significant confusion noise, appropriate for the poorer u-v coverage at the time scales we are investigating.

\begin{figure}
    \centering
    \includegraphics[angle=0,width=0.45\textwidth]{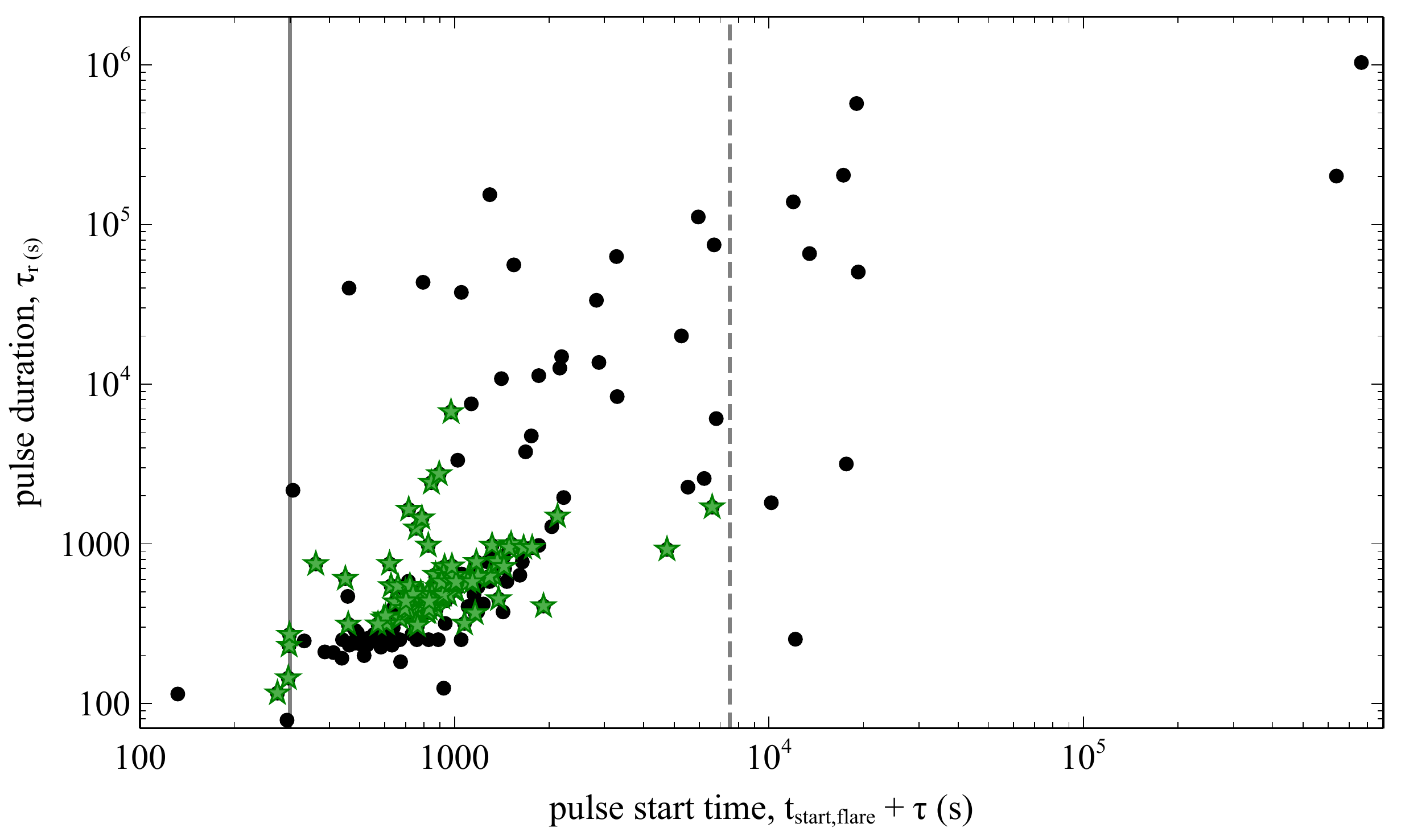}
    \caption{Pulse duration against pulse start time (observer frame). Green stars indicate the 84 flares identified as detectable in Figure \ref{fig:144mhz_flux_duration}. The grey line indicates the current LOFAR response time, and the grey dashed line indicates 2 hours later, representing a typical integration.}
    \label{fig:144mhz_duration_delay}
\end{figure}

The flares have a range of durations, $\tau_r$, and delays, $\tau$, from the start of the observation (Figure \ref{fig:144mhz_delayhist}). We need to determine whether the flares in Figure \ref{fig:144mhz_flux_duration} are detectable for a given trigger observation, and for this we will work under the assumption that flares have a constant flux density top-hat structure. The telescope has a response time, $T_{\rm resp}$, to transient events and for LOFAR the guaranteed response time is 5 minutes. Then the observation has a given duration, $T_{\rm dur}$, which we take to be 2 hours in this analysis. This leads to two scenarios:
 
\begin{itemize}
	\item $\tau + \tau_r \le T_{\rm resp} + T_{\rm dur}$: in which the end of the predicted radio pulse is shorter than the duration of the observation. The detectability of the radio pulse is then determined by its duration, its flux density and the sensitivity of the telescope for the duration of the pulse (as in Figure \ref{fig:144mhz_flux_duration}).
	\item $\tau + \tau_r > T_{\rm resp} + T_{\rm dur}$: in this case, the radio pulse is much longer than the observation. Therefore, we need to determine if the pulse would be detectable during the telescope observation window. The observable duration, $T_{\rm obs}$, is given by: $T_{\rm obs} = \tau_r - [(\tau + \tau_r) - (T_{\rm resp} +T_{\rm dur})]$. We can then predict the sensitivity of the radio observation for an integration time equal to $T_{\rm obs}$ (using Equation \ref{eqn:rmsnoise}). If the predicted flux density of the radio flare is greater than the expected sensitivity in the image, then the radio flare is detectable. 
\end{itemize}
Performing this calculation, we find 3 flares among the detected subset with stop times after the observation is complete, and of these, two remain detectable. We can apply the same logic to flares beginning before the 5 minutes response time but with durations that fall inside the observing window. There are 6 of these flares, among which 4 would be detectable.

The resulting proportion of this X-ray flare sample detectable by LOFAR in rapid response mode is 44 per cent.

\section{Discussion}

\subsection{Summary of results}
We find that a significant fraction, 44 per cent, of X-ray flares should be detectable at low frequencies with current LOFAR instrumentation, if the prompt emission is produced via magnetically dominated processes as described in \citet{Usov2000}. This is very promising for observational searches. Predictions for observing radio flares from short GRBs have been examined using this same model, focusing on the prompt $\gamma$-ray flares, and have been shown to be observable with the fastest response radio facilities \citep{Rowlinson2019}. 

\subsection{X-ray flares as prompt emission} 
We present evidence in Section \ref{sec:intro} for the connection between prompt $\gamma$-ray pulses and X-ray flares, which is a premise of this analysis, and we now examine this for our sample of flares in particular.
Among our 200-strong flare sample, 71 may have started whilst the prompt $\gamma$-ray emission was ongoing, since the flare start time lies within the GRB duration according to BAT $T_{\rm 90}$.
On the other hand, some flares are really late even in the rest frame and, while it is difficult to assign these to the same internal mechanism at least one late flare study showed that flares occurring $>10^4$ s after trigger (observed frame), generally have the same properties as the larger early flare sample \citep{Curran2008}. These very late flares cannot be probed with 2 hr integration times however, so would not be targeted for LOFAR observations.

\citet{Oganesyan2018} studied GRBs with overlapping {\it Swift} BAT-XRT coverage and fitted their time-averaged joint spectra, including also {\it Fermi} GBM data where possible. Their sample spans the time frame of the \citet{Yi2016} flare sample, and includes 15 of the GRBs considered in this work (all long, 5 with GBM data). Their results showed that 11 are best-fitted with a Band or Band cut function, with a low energy break in the XRT band, 1 with a cut-off power law, and 3 sources can be described by a single power law.
They apply an offset correction factor between instruments, but in no cases were these too large to accommodate a single origin for the X-ray to $\gamma$-ray emission. 
\citet{Oganesyan2018} argue that even when a low energy break is required, the spectrum can be considered a single, prompt synchrotron component in a fast-cooling regime.

\subsection{LOFAR rate predictions}
We note that this flare sample probes only a fraction of the total GRB sample which could be accessed by rapid response triggered observations, since it has been limited to known redshift sources (this information usually becomes available only at later times). We therefore estimate the rate of detectable flares with LOFAR, $R_{\rm LOFAR}$, starting from the {\it Swift} GRB rate of $R_{\rm GRBs}$ of 88.7 yr$^{-1}$ (averaged over 2005--2019), 
\begin{equation} \label{eqn:ratecalc2}
R_{\rm LOFAR} =  R_{\rm GRBs} \times f_{\rm XRT} \times f_{\rm flare} \times f_{\rm sky} \times f_{\rm detec} = 7.1~{\rm yr}^{-1}   
\end{equation}
where the fraction of GRBs detected with XRT, $f_{XRT}$, is 95\% \citep{Evans2009}, the fraction of GRBs with X-ray flares, $f_{\rm flare}$, is taken as 48 per cent \citep{Swenson2014}, the detectable fraction to LOFAR in rapid response mode at 144 MHz, $f_{\rm detec}$, is 44 per cent from this work, and assuming 40 per cent sky coverage, $f_{\rm sky}$. Just over 20 per cent of LOFAR-accessible GRBs are expected to have an X-ray flare that would be detectable at $\ge 5\sigma$. This can be optimised given the direct correspondence in the adopted model between $\gamma$-ray fluence and radio flux density. A focus on the more energetic GRBs, for example by excluding image triggers (which comprise 17.5\% of {\it Swift} triggers, \citealt{Lien2016}) and low significance rate triggers, could achieve this. 

A byproduct of this selection is that bright GRBs are much more likely to have prompt detections with the UVOT instrument and other optical telescopes, leading to a greater chance of redshift determination and tighter constraints on the model parameters. Among all {\it Swift} GRBs only 31 per cent have prompt detections with the UVOT instrument, while for the flare sample used here this is 66 per cent.

\subsection{Future prospects}
Detection of coherent radio emission during the prompt emission phase of a GRB would be a strong indication that a Poynting flux mechanism is preferred over the matter-dominated fireball and other similar models. Detections, or constraining limits, for a small sample of GRBs would allow this result to be generalised and inform further development of prompt emission models. 

Work has started to upgrade LOFAR to LOFAR2.0\footnote{\url{https://www.astron.nl/nl/eu-funded-research/lofar-2-20s}}, a significant upgrade that will lead to future improvements for this project. Distinction between the prompt and afterglow phases is important, hence the desire for earlier radio detections. Decreasing the response time may be a realistic prospect for LOFAR2.0. Pushing the LOFAR response time back from 5 minutes to 4.5 minutes, the number of detectable flares fully within the time window that we predict from our study would increase from 85 to 89, but no further partial flares would be detected. This increases the detected fraction by just 1\%, but we acknowledge an increasing incompleteness in the X-ray flare sample at earlier times dependent upon XRT slew time (Section \ref{sec:sample}). Additionally, one of the key goals of LOFAR2.0 is to significantly enhance the LBA capabilities, leading to more sensitive images in the lowest frequency band and enabling the simultaneous use of the HBA and LBA modes. This will enable us to place tighter constraints at the lower radio frequencies where this emission is expected to peak.

With the recent enhancements to the Murchison Widefield Array \citep[MWA;][]{tingay2013} and its significantly faster rapid response mode with slew times as short as 8 s \citep{hancock2019}, some of the earliest X-ray flares and prompt $\gamma$-ray flares may be detectable providing complimentary observations to those obtained by LOFAR \citep[note MWA is less sensitive than LOFAR for longer integrations;][]{Rowlinson2019}. Additionally, in the near future, the Square Kilometer Array \citep[SKA;][]{dewdney2009}\footnote{\url{https://www.skatelescope.org}} will be built and comprises two facilities: SKA-Mid (350 MHz -- 14 GHz) and SKA-Low (50 -- 350 MHz). SKA-Low is expected to provide approximately an order of magnitude improvement in image sensitivity compared to LOFAR and a faster response time, leading to a significantly larger population of radio flares predicted to be observed with the observed $\gamma$-ray and X-ray flares.

The launch of the Space Variable Objects Monitor satellite \citep[SVOM,][]{Wei2016}, planned for launch end 2021, will provide prompt X-ray data down to 4\,keV using the ECLAIRS coded mask instrument, with which $\sim$70 GRB detections per year are expected. This means that X-ray flares can be picked up at much earlier times, and with GRB localisations in the arcminutes range adequate for immediate LOFAR pointings. 

\section{Conclusions}
We put forward an observational test of the magnetically-dominated wind model of \citet{Usov2000} as the mechanism for GRB prompt emission. This model predicts low-frequency radio pulses associated with the prompt pulses, which peak at MHz frequencies and may be probed with LOFAR in rapid response mode. We use X-ray flares, shown to arise from sites internal to the jet, to probe GRB prompt emission out to later times, easily accessible with LOFAR. Adopting the {\it Swift} XRT flare sample compiled by \citet{Yi2016} we apply a magnetically-dominated wind model to make predictions for the timing and flux density of corresponding radio pulses. We find that 44 per cent of the flares in this sample would have been detectable with LOFAR HBA at 144 MHz under this model, for typical sensitivities reached using previously executed rapid response mode observations \citep{Rowlinson2019b} and assuming negligible absorption and scattering effects in the interstellar and intergalactic medium. We estimate a rate of order seven {\it Swift} GRBs per year that would be both accessible to and detected during X-ray flaring by LOFAR. We conclude that such triggered low-frequency radio observations can play a key role in establishing the long debated mechanism, either magnetically- or matter-dominated, responsible for GRB prompt emission.

\section*{Acknowledgements}
This work was made possible by the ASTRON Helena Kluyver visitor programme enabling RLCS to make an extended visit to AR. RLCS also acknowledges funding from STFC, and is indebted to L. and K. Wiersema for their support.
This work made use of data supplied by the UK {\it Swift} Science Data centre at the University of Leicester. 
This paper is based (in part) on data obtained with the International LOFAR Telescope (ILT). LOFAR is the Low Frequency Array designed and constructed by ASTRON. The authors thank S. ter Veen for useful discussions and J. Katz for a constructive review.








\appendix


\bsp	
\label{lastpage}
\end{document}